\documentclass[12pt]{article}
\raggedright   %
\usepackage{amssymb}
\usepackage{latexsym}
\newfont{\tnbf}{cmbx10}
\newfont{\ninebf}{cmbx9}
\def\ket#1{|\,#1\,\rangle}

\begin{document}
\baselineskip=14pt
\begin{center}
{\large\bf A Possible Explanation of the Superposition Principle}
\\
\bigskip\bigskip
{\sl Kent A.\ Peacock \\
Department of Philosophy \\
University of Lethbridge.}\footnote{Email:  kent.peacock@uleth.ca}
\end{center}
\begin{quote}
  {\bf ABSTRACT:}  I tentatively suggest that the superposition principle of quantum mechanics
is explicable in a mathematically natural way if it is possible to understand
probability amplitudes as complex-valued logarithms.  This notion is
inspired by the fact that the quantum state may be interpreted as a
measure of information.
\end{quote}
The object of this note is to sketch what J.\ A.\ Wheeler might
call an ``idea for an idea'' \cite{MWT}.  The notion I present
here points the way toward a possible new interpretation of
quantum mechanics, although the development of this approach that
I am able to offer in this paper is in itself certainly not yet
complete enough to deserve such a lofty description.  I hope to
produce a more thorough treatment soon; but in the meantime I dare
to think that the bare notion I describe here may be of sufficient
interest to merit an airing in preprint form, and I commend it to
the attention of those who may be more mathematically skilled than
I am.

\subsection*{I.  The Feynman Problem} Quantum mechanics is beset by a
number of interpretational challenges arising from spectacularly
non-classical phenomena such as nonlocality and superfluidity, as
well as the host of difficulties surrounding the measurement
problem.  But the deepest mysteries surround our lack of
understanding of origin of the basic rules of the theory.

We can frame the problem very clearly by going back to the
statement of the basic principles of quantum mechanics given by
Richard Feynman and his co-authors in {\sl The Feynman Lectures on
Physics} \cite{FLS64}:
\vskip16pt
{\small   
\noindent{(1)  The probability of an event in an ideal experiment is given
by the square of the absolute value of a complex number $\phi$
which is called the probability amplitude:
\begin{eqnarray}
  P & = & \mathrm{probability} \\
  \phi & = & \mathrm{probability\; amplitude}  \\
  P & = & |\phi|^2.
\end{eqnarray}
}

\noindent{(2) When an event can occur in several alternative ways, the
probability amplitude for the event is the sum of the probability
amplitudes for each way considered separately.  There is
interference:
\begin{eqnarray}
   \phi & = & \phi_1 + \phi_2, \\
   P & = & |\phi_1 + \phi_2|^2.
\end{eqnarray}
}

\noindent{(3)  If an experiment is performed which is capable of
determining whether one or another alternative is actually taken,
the probability of the event is the sum of the probabilities for
each alternative.  The interference in lost:
\begin{equation}
   P =  P_1 + P_2.
\end{equation}
}
\cite[p. 1--10]{FLS64}  [I have renumbered the equations.]
}  
\vskip16pt

Eq.\ 3 is, of course, also known as the Born Rule.  As it shows,
the probability amplitude is, so to speak, the ``square root'' of
the probability.

All the multifarious and complex developments of quantum mechanics
are applications of these rules, which, so far, must simply be
taken for granted.  As Feynman {\it et al.} say,

\vskip16pt {\noindent\small One might still like to ask:  ``How
does it work?  What is the machinery behind the law?''  No one has
found any machinery behind the law.  No one can ``explain'' any
more than we have just ``explained''.  No one will give you any
deeper representation of the situation.  We have no ideas about a
more basic mechanism from which these results can be deduced.
\cite[p. 1--10]{FLS64}  [See also similar remarks in
\cite{Feynman65}.]
}
\vskip16pt

I will call the problem of explaining rules (1) to (3) the
``Feynman Problem''.   It seems frustrating that something as
simple and basic as these rules cannot be explained. This
uncomfortable fact implies that we still do not understand the
deepest principles of physics, despite the considerable power and
sophistication that our physical science has already attained.

There is a clarification that can immediately be added to the
above statement of quantum principles.    As Feynman {\it et al.}
go on to explain, what point (2) really says is that if there is
no way of telling which route the system takes {\em without
measuring a non-commuting observable}, then interference of
amplitudes may occur.  The classic illustration of this
(\cite{FLS64}) is the double-slit experiment:  if we do something
that allows us to determine which slit the electrons go through
(thereby finding their positions), we wipe out the interference
pattern (which depends upon the particles being in pure momentum
states).  There would be no interference if all observables
commuted. Since all uniquely quantum phenomena are in some way a
consequence of interference, all uniquely quantum phenomena are
somehow a consequence of non-commutativity. (For a recent and
forceful expression of this view, see \cite{Bub97}.) Therefore, we
could add to Feynman's list the fundamental problems of
understanding the origin of non-commutativity and the intimately
related problem of understanding the origin of Planck's constant
of action.

The problem, then, is to answer the following inter-related
questions:
\begin{enumerate}
\item{Why do we have to represent quantum processes and states by
  {\em complex-valued} mathematical objects?}
\item{Why the superposition principle? --- That is, why do we represent
 quantum states and processes by objects that add up linearly?}
\item{Where does the Born rule come from?  Why is this the right
  way to calculate probabilities?}
\item{Why non-commutativity?}
\item{Why does Planck's constant of action have the particular
  magnitude that it has?}
\end{enumerate}

(Questions 1 and 2 can be combined into the question, ``Why
Hilbert space?'')

In the following I will offer a tentative answer to the second
question, and make some hesitant suggestions about the others.

\subsection*{II.  The Quantum State as a Measure of Information} In
recent years, a very fruitful interpretation of the quantum state
has begun to emerge --- the notion that state vectors are some
sort of measure of {\em information}  \cite{NC2000}. Indeed, one
often speaks of state vectors of the form
\begin{equation}
\ket{\psi} = \alpha\ket{0} + \beta\ket{1}
\end{equation}
as ``qubits'', in analogy to the bits (``binary information
units'') of the classical information theory developed by Shannon
and others \cite{SW48}.  (The apt term ``qubit'' is due to
Schumacher \cite{Schu95}.)  In this formula, $\ket{0}$ and
$\ket{1}$ are interpreted most naturally as eigenstates of some
Hermitian observable (such as the spin of an electron) that is
binary in the sense that its spectrum of eigenvalues is the set
$\{0, 1\}$.  The complex numbers $\alpha$ and $\beta$ are phase
factors such that
\begin{eqnarray}
  \mathrm{Prob}(\mathrm{getting\; 0}) & = & \alpha^2,   \\
  \mathrm{Prob}(\mathrm{getting\; 1}) & = & \beta^2, \;\;{\mathrm{and}} \\
  \alpha^2 + \beta^2 = 1.
\end{eqnarray}

In the past ten years or so we therefore see the idea emerging
that quantum mechanics is a sort of generalized information
theory.  As Gerard Milburn puts it,

\vskip16pt
{\small \noindent\dots quantum theory, our best theory
of physical
  reality, is
  actually a theory, not of physical {\em things}, but of physical
  {\em information} (even today not every physicist would accept
  this point of view).  \cite[p.\ 153]{Milburn97}
}
\vskip16pt

\subsection*{III.  Taking the Information-theoretic Interpretation
  Seriously}
If we are to take seriously the notion that quantum
theory is a generalization of classical information theory, then
we should push the analogy between quantum and classical
information theory as far as possible.  In the classical theory of
Shannon, information is a {\em logarithm} --- the log of the
complexity (or multiplicity) of a system; that is, the log of the
number of ways that a process could have gone.   We use a
logarithmic measure of complexity for the convenient reason that
logarithms are additive. The multiplicity of a number of
experiments done in succession is the product of their individual
complexities; equivalently, the information required to express
the result of a concatenated series of experiments is the sum of
the information to be had from each individual experiment.

To make the process clear, consider the elementary example of a
two-state system such as a coin toss.  A single toss can come out
in two ways, but its result can be represented by one letter, H or
T.  Two tosses in succession can come out four different ways, but
we only need two letters to represent the outcome; $n$ tosses can
come out in $2^n$ ways, but we only need $n$ letters to represent
the outcome, and $n$ tosses followed by $m$ tosses can be
described by $n + m$ letters.  Shannon therefore found it
convenient to define the information contained in the outcomes of
$n$ binary events by the binary logarithm of the complexity. This
can be generalized easily to experiments in which the outcomes are
not equiprobable. The key point, again, is that the simplicity
comes from the additivity of logarithms.

My suggestion is that the linearity of quantum mechanics might
naturally be explained if we could show that {\em probability
amplitudes can be treated as logarithms}.  (Following Feynman's
approach, we focus our attention on the probability amplitude;
this involves no loss of generality, since state vectors such as
Schumacher's qubits are simply arrays of amplitudes.) These logs
must be complex-valued, and this raises problems of interpretation
which I will discuss here but not definitively solve.  But suppose
that we can intuitively think of amplitudes as representing the
logarithms of a complexity, however that complexity may be defined
precisely. The superposition of amplitudes would then correspond
to the multiplication of the complexities associated with those
amplitudes. If this can be made to work in detail, then we would
see that quantum mechanics is essentially a calculus of
complex-valued logarithms.

Again, here is my conjecture:
\begin{quote}
  Probability amplitudes are complex-valued logarithms of a
  complex-valued complexity associated (in a way to be
  determined) with physical transitions of state.  The
  superposition of probability amplitudes (leading to non-classical
  interference phenomena because the amplitudes are
  complex) corresponds to the multiplication of complexities
  associated with the processes associated with the amplitudes
  for those states.
\end{quote}

The idea outlined here is, as emphasized, highly tentative and
sketchy. However, it is the first notion that I have encountered
that could even count as a candidate for an explanation for the
linearity of quantum mechanics, and it needs to be investigated
with some care.

\subsection*{IV.  Concluding Observations and Speculations}
The notion of
seeking a logarithmic interpretation of the elements of a linear
vector space is not as odd as it might seem at first glance.
Consider, for instance, the set of the first $n$ primes, including
1, and their inverses.  Take the logs (to any convenient base) of
the members of this set.  The log of any composite number that is
built up out of multiples of the given $n$ primes (but no others)
will be a linear combination of the logs of the $n$ primes.  Logs
of the reciprocals give additive inverses, and $\log{1}$ gives the
zero vector.  Each such set of $n$ primes together with their
inverses therefore defines a vector space of dimension $n-1$, with
the logs of the composite numbers acting as vectors in the space.
Whether this has any useful application to number theory I am not
sure, but it raises the interesting question of determining the
conditions under which any vector space can be usefully
interpreted as a space of logarithms.  In particular, this raises
the disturbing possibility, which I mention only in passing, that
ordinary 3-d position space, or space-time itself, could
conceivably accept such an interpretation.

The notion that we could explain the superposition principle if
probability amplitudes are a kind of logarithm certainly seems
mathematically natural.  However, to make this workable we still
need to understand what it could mean to talk about complex-valued
complexities.  We also need to see where the Born Rule comes from.
Why can we get back to a classical probability by taking the
modulus of a complex logarithm?

I have no clear idea at this writing what the answer to the second
question would be, although I suspect that it will turn out to be
mathematically obvious.  The first question is deeper, but one can
see a direction that could be worth exploring.

The key may be to follow up Feynman's pregnant suggestions about
negative probabilities \cite{Feynman87}.  Feynman pointed out that
negative probabilities are just as sensible, and possibly just as
useful, as ordinary negative numbers, so long as they are used
only in intermediate calculations whose final results come out
positive:  ``\dots conditional probabilities and probabilities of
imagined intermediary states may be negative in a calculation of
probabilities of physical events or states'' \cite[p.\
238]{Feynman87}.  And, as Feynman explains, one situation in which
we could expect negative probabilities to arise naturally is when
a system may be in one of two mutually incompatible conditions,
such as a quantum system that may be subjected to non-commuting
measurement procedures.

\vskip16pt
{\small
\noindent  If a physical theory for calculating probabilities yields a
  negative probability for a given situation under certain assumed
  conditions, we need not conclude that the theory is incorrect.
  Two other possibilities of interpretation exist.  One is that
  the conditions (for example, initial conditions) may not be
  capable of being realized in the physical world.  The other
  possibility is that the situation for which the probability
  appears to be negative is not one that can be verified directly.
  A combination of these two, limitation of verifiability and
  freedom in initial conditions, may also be a solution to the
  apparent difficulty.  \cite[p.\ 238--9]{Feynman87}
}
\vskip16pt

Now, it is a very short step from negative frequencies (or
probabilities) to complex-valued ``roots'' of frequencies (or
probabilities)
--- if for no other reason than the fact that if negative
probabilities obey algebraic relations higher than first degree,
the fundamental theorem of algebra guarantees that these relations
may well be satisfied in some cases by complex-valued quantities.
My point, therefore, is that there may be a deep but natural
connection between the use of complex numbers to represent, as it
were, ``square roots'' of probabilities, and the fact of
non-commutativity.

Is the difference between classical and quantum information {\em
merely} the move from real-valued to complex-valued measures of
information?  Non-commuting observables $\hat{A}$ and $\hat{B}$ in
quantum mechanics obey commutation relations of the general form
\begin{equation}
  [\hat{A}, \hat{B}] = i\hbar \hat{C}
\end{equation}
where $\hbar$ is Planck's reduced constant.  For all we know,
$\hbar$ might have a different numerical value than it happens to
have; and therefore there might be a whole class of mathematically
possible quantum information theories depending on the value of
$\hbar$.  On the other hand, it may be that $\hbar$ is somehow
mathematically determined, in which case there is only one
mathematically possible quantum mechanics.  However, I am rapidly
approaching the limits of permissible speculation here, and I will
conclude by reiterating the point that the need for
complex-valued frequencies or complexities
is very likely related to the fact of
non-commutativity, in a way that remains to be made clear.

If anything like what I suggest here is right, the linearity of
quantum theory is a mathematical artifact, stemming from the use
of a logarithmic description.  It is quite possible, and rather
 important to note, that the underlying
{\em dynamics} of quantum phenomena might be highly nonlinear
--- and indeed this is suggested by the fact of non-commutativity.

It is beyond the scope of this paper to compare the  theory
sketched here with the numerous other interpretations of state
vectors or wave functions that have been advanced in the past. At
first glance, it does seem to conflict with realistic
interpretations of the wave function such as Everett's relative
state formulation \cite{MWT} or Bohm's interpretation \cite{BH93}.
On the other hand, a tsunami can be thought of as merely a
probability distribution of water molecules, but it can still pack
quite a punch.  It would be well to avoid commitment to
philosophical preconceptions about the meaning of the quantum
state until we are much clearer about the mathematical
possibilities---for a likely implication of recent trends in
quantum information theory is that there is much more to quantum
mechanics that is {\em purely mathematical} than first meets the
eye.

\subsection*{Acknowledgements} The notion that we could account for
the superposition principle if we think of the quantum state as a
logarithm occurred to me during conversation with John L.\ Bell. I
am grateful to him for many inspiring conversations, although he
is, of course, not responsible for any errors of my own.   Many
thanks to the University of Lethbridge and to the Social Sciences
and Humanities Research Council of Canada for financial support.


{
}

\end{document}